\documentclass[%
 reprint,
 amsmath,amssymb,
 aps,
]{revtex4-2}
\usepackage[utf8]{inputenc}
\usepackage{graphicx}
\usepackage{dcolumn}
\usepackage{bm}

\begin{document}

\preprint{APS/123-QED}

\title{Electrically driven singlet-triplet transition in triangulene spin-1 chains}

\author{Gabriel Martínez-Carracedo$^{1,2}$}
\author{László Oroszlány$^{3,4}$}
\author{Amador García-Fuente$^{1,2}$}
\author{László Szunyogh$^{5,6}$}
\author{Jaime Ferrer$^{1,2}$}

\affiliation{
$^1$Departamento de Física,  Universidad de Oviedo,  33007 Oviedo, Spain\\
$^2$Centro de Investigación en Nanomateriales y Nanotecnología, Universidad de Oviedo-CSIC, 33940, El Entrego, Spain\\
$^3$Department of Physics of Complex Systems, Eötvös  Loránd University, 1117 Budapest, Hungary\\
$^4$MTA-BME Lendület Topology and Correlation Research Group, Budapest University of Technology and Economics, 1521 Budapest, Hungary\\
$^5$Department of Theoretical Physics, Institute of Physics, Budapest University of Technology and Economics, M\H{u}egyetem rkp. 3., H-1111 Budapest, Hungary \\
$^6$ELKH-BME Condensed Matter Research Group, Budapest University of Technology and Economics, M\H{u}egyetem rkp. 3., H-1111 Budapest, Hungary
\looseness=-1}

\begin{abstract}
Recently, graphene triangulene chains have been synthesized and their magnetic response has been analyzed by STM
methods by Mishra and coworkers (\textit{Nature} {\bf{598}}, 287 (2021)). Motivated by this study,
we determine the exchange bilinear 
and biquadratic constants of the triangulene chains by calculating two-spin rotations in the spirit of the magnetic
force theorem. We then 
analyze open-ended, odd-numbered chains, whose edge states pair up forming a triplet ground state. We propose three 
experimental approaches that enable us to trigger and control a singlet-triplet spin transition. Two of these 
methods are based on applying a mechanical distortion to the chain. 
We finally show that the transition can be controlled efficiently by the application of an electric field. 
\end{abstract}

\maketitle


\section{Introduction}
Simple spin models have played a key role in the formulation and comprehension of 
the basic principles of magnetism and statistical mechanics since the early days of quantum theory
\cite{ising25,heisenberg28}. The interest in these models and in the systems realizing them 
persists today due to their connection to many topological properties of matter \cite{haldane83,kitaev03},
as well as their potential to become the building blocks of viable and robust quantum computers \cite{wei12,kitaev06}.
Infinite quantum antiferromagnetic (AFM) spin-1 chains have a singlet ground state and a gap in their
excitation spectrum \cite{haldane83}. This is because each atomic spin fractionalizes into two spin-1/2 states, and
each spin-1/2 state entangles with another one at a neighbor site forming a singlet. The spin-1 chain therefore
decomposes into a set of singlet dimers \cite{AKLT87,auerbachbook}. Further numerical 
work \cite{Kennedy90,white93,auerbachbook} on the open-ended bilinear-biquadratic (BLBQ) nearest neighbor model 
\begin{equation}
\hat{H}_{BLBQ}=J\sum\limits_{i=1}^{N-1}\,
\left[
\mathbf{\hat{S}}_i\cdot \mathbf{\hat{S}}_{i+1}+\beta\left(\mathbf{\hat{S}}_i\cdot \mathbf{\hat{S}}_{i+1} \right)^2
\right]
\label{BLBQ}
\end{equation}
has shown that a spin-1/2 edge excitation appears at each of the two ends, whose energy lies inside the Haldane gap.
These edge states entangle into a singlet and a triplet, whereby the singlet/triplet is the ground state for 
even/odd $N$-chains, and the singlet-triplet energy splitting $\Delta E_{ST}=E(S=1)-E(S=0)$ decays exponentially 
with the chain length. 
\begin{figure}[h]
    \centering
    \includegraphics[width=\columnwidth]{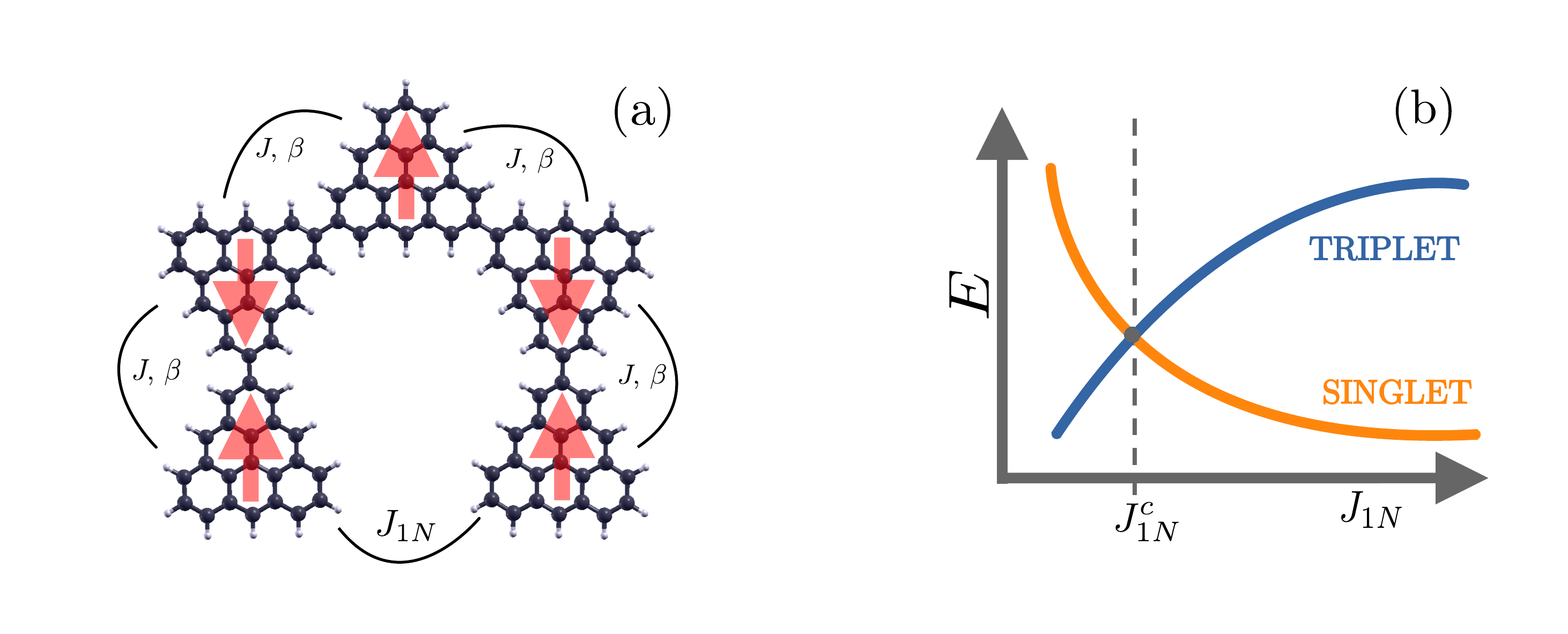}
    \caption{(a) Sketch of a horseshoe-like $N=5$ GT chain, where the spins at each triangulene pair are coupled by the 
    same exchange constants $J$ and $\beta$, and where the two end GT spins are coupled by a smaller exchange 
    constant $J_{1N}$. (b) A singlet-triplet crossing occurs at a finite value of $J_{1N}/J$ 
    smaller than 1.}
    \label{Figure: motivation}
\end{figure}
Theoretical and experimental works have explored already the potential of the singlet/triplet transition of these 
edge modes for storage and manipulation of quantum information \cite{wei12,shulman12,jaworowski17,sompet22}.
However, efforts to realize unequivocally quantum spin-1 chains have been hindered by a variety of factors among which the 
magnetic anisotropy arising from the spin-orbit interaction is possible the most relevant one. Recently, graphene 
triangulenes (GT) have 
been synthesized as single molecules \cite{Su20, wang22}, or forming chains \cite{mishra21}, where due to Lieb's theorem
\cite{lieb89,rossier07} each triangulene block is characterized by a robust  spin-1 magnetic moment. Because the constituent 
carbon atoms have a negligible spin-orbit interaction, GT chains are faithful realizations of the open-ended spin-1 
quantum AFM chain model embodied in Eq. (\ref{BLBQ}).

We propose in this article three experimental approaches to trigger and control a singlet-triplet transition for 
odd-numbered AFM spin-1 GT chains. The proposals are based on the experimental bottom-up approach of Ref.  
\citet{mishra21} that leads to chains of many different lengths and shapes. Specifically, horseshoe-shaped 
chains of different lengths were synthesized, see the example sketched in Fig. \ref{Figure: motivation} (a). 
Because increasing the length $N$ of the chain increases its ductility, the two ends of the chain can be
brought in close proximity, which in turn introduces an exchange coupling $J_{1N}$ between the two magnetic degrees
of freedom localized at the edges. Since the ground state of odd-numbered chains is a triplet, while that of a cyclic
chain is a singlet, there must be a critical $J_{1N}^c$ separating the two ground states, as depicted in Fig. 1(b).

The experimental feasibility of the proposals is ensured by the ability to manipulate and measure spectroscopically 
graphene nanostructures by Scanning Tunneling Microscopy (STM) methods \cite{mishra21,fasel16,oteyza20,oteyza22}.
We establish first the requirements for a triplet-singlet level crossing via exact diagonalization. We then use
a first principles approach to map GT chains to one-dimensional spin-1 Heisenberg chains and extract the corresponding 
exchange constants. The main prediction of our study is that the critical inter-edge constant $J_{1N}^c$ can be reached 
by experimentally feasible mechanisms, especially, by the application of an external electric field.
\section{Method}
\subsection{Exact Diagonalization}
We compute here the energy spectrum of the Hamiltonian
\begin{equation}
\hat{H}=\hat{H}_{BLBQ}+J_{1N}\,\mathbf{\hat{S}}_{N}\cdot \mathbf{\hat{S}}_{1}
\end{equation}  
for odd-{\it N} chains from  $N\,=\,3$ to $15$.  Our calculations for $J_{1N}=0$ show that a singlet 
and a triplet edge states lie inside the Haldane gap as expected, the triplet being the ground state.  
 
\begin{figure}
    \centering
    \includegraphics[width=\columnwidth]{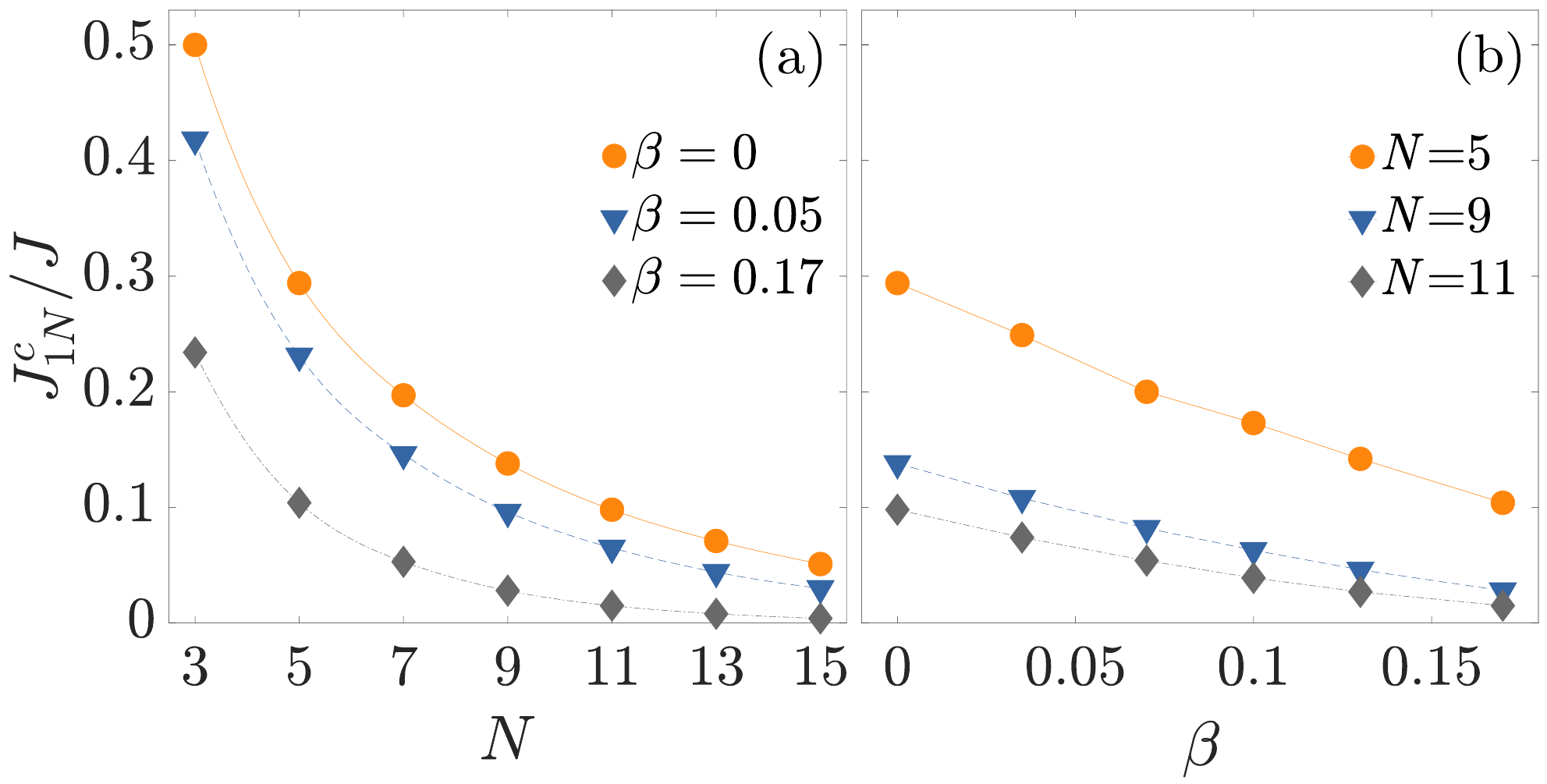}
    \caption{(a) $J_{1N}^c/J$ as a function of $N$ for several values of $\beta$. (b) $J_{1N}^c/J$ as a function 
    of $\beta$ for chains having different lengths. }
    \label{Jcritic}
\end{figure}

We search for the critical value $J_{1N}^c$ of the exchange constant, that renders a four-fold degenerate
ground state.  Fig. \ref{Jcritic}(a) shows that $J_{1N}^c/J$ decays exponentially 
with $N$.  Fig. \ref{Jcritic}(b) demonstrates that larger values of $\beta$ facilitate reaching the critical 
$J_{1N}^c$.  All in all, we find that the singlet-triplet crossing happens at reasonably small values of 
$J_{1N}/J \sim 0.01 - 0.2$ if $N$ is larger than 7,  and for values of the biquadratic parameter 
$\beta$ relevant for the GT chains extracted both experimentally \cite{mishra21} and in the first principles 
mapping presented below.

The exponential decay of $J_{1N}^c$ with $N$ means that the singlet-triplet energy splitting
$\Delta E_{ST}$ also decays with $N$.  We find that values of
$\Delta E_{ST}$ larger than about 1 meV require exchange constants $J_{1N}/J>0.1$ (calculations with many values of $N$ are shown at \cite{supmat}).   
We show in the next section that the energy scale of the singlet/triplet splitting can be tuned in the order
 of a few meV, so that it should be easily resolved experimentally using spectroscopic methods. 
 
\subsection{Ab initio simulations of GT chains}
We have carried out Density Functional Theory (DFT) simulations of GT chains having an odd number of 
GTs, as shown in Fig.   \ref{Figure: motivation} (a).  Each GT contains 22 carbon atoms, and has zigzag edges 
where each edge carbon atom has been passivated with hydrogen.  We have used the DFT package SIESTA \cite{siesta},
with the Generalized Gradient Approximation \cite{PBE}.  We have used established pseudopotentials for carbon and
hydrogen,  and strict accuracy tolerances such as a real-space grid cutoff of 500 Ry. We have first confirmed 
that the 
total spin of a single isolated GT is $s=S/\hbar=1$ via a Mulliken analysis. We have then simulated odd-numbered 
chains possessing AFM spin alignment. We have found that the total charge and spin of each GT in the chain are
the same as those of an isolated GT up to four-five decimal digits. We therefore conclude that charge fluctuations 
among GTs are frozen, and that the low-energy sector of the Hilbert space of each GT corresponds to that of a
quantum spin-1 degree of freedom.
\subsection{Mapping to the BLBQ model}
We can extract the bilinear ($J$) and biquadratic ($\beta$) constants that couple the spin-1 degrees associated to GTs,
by making use of the fact that in absence of spin-orbit coupling any collinear state is either stable or metastable. 
The energy cost of infinitesimal rotations of the spins from their collinear reference states at two 
different GTs ($n \ne m$) in a chain can be expanded to second order as \cite{supmat}
\begin{equation}
    \delta E_{nm}^{(2)}=D_{nm}^{(2)}\,
    \delta\mathbf{S}_n\cdot\delta\mathbf{S}_m \,
    \label{Eint}
\end{equation}
where
\begin{equation}
    D_{nm}^{(2)}=J_{nm}\,\left(\,1+2\,\beta_{nm}\left(\mathbf{S}_n\cdot\mathbf{S}_m\right)\,\right)\, .
    \label{Dnm}
\end{equation}
We apply the generalization of the LKAG formula to the case of a non-orthonormal basis set \cite{oroszlany19} to determine 
$D_{nm}^{(2)}$, in the spirit of the magnetic force theorem \cite{liechtenstein87,katsnelson00,udvardi03}. 
We compute $D_{nm}^{(2)}$ for both the FM and AFM reference spin configurations to solve for $J_{nm}$ and $\beta_{nm}$. 
This yields
\begin{eqnarray}
J_{nm}     & = & \dfrac{1}{2}\,\left(D_{nm}^{(2),FM}+ D_{nm}^{(2),AFM}\right)\\
\beta_{nm} & = & \dfrac{1}{2}\,\dfrac{D_{nm}^{(2),FM}- D_{nm}^{(2),AFM}}{D_{nm}^{(2),FM}+D_{nm}^{(2),AFM}} \, .
    \label{Jbeta}
\end{eqnarray}

Our results for the nearest-neighbor constants of an infinite GT chain and a GT dimer are shown in Table~\ref{tableJandB}.
We have also written in the Table the values obtained experimentally in Ref. \cite{mishra21}, where STM data were used 
to fit the spectrum of Eq. \eqref{BLBQ}. Albeit our procedure gives somewhat higher values for both $J$ and $\beta$, the 
agreement between our parameter-free first-principles approach and the experimental fittings of 
Ref. \cite{mishra21} is remarkable.

Our method allows us to determine the exchange constants between any two GT sites $n$ and $m$ in the chain. We have 
therefore computed the next-nearest neighbor constants $J_{n\,n+2}$ and $\beta_{n\,n+2}$ and found that they are three to 
four orders of magnitude smaller than the nearest-neighbor parameters $J$ and $\beta$, providing compelling evidence of 
the accurate realization of open-ended and cyclic nearest-neighbor quantum AFM chains by the GT chains synthesized 
in Ref. \cite{mishra21}.

\begin{table}[h]
\small
  \caption{\label{tableJandB}
Nearest-neighbor $J$ and $\beta$ constants for a GT dimer, for an infinite GT chain obtained from Eqs. (\ref{Jbeta}) and
from the fit to STM experiments performed in Ref. \cite{mishra21}.}
\label{tbl:example}
  \begin{tabular*}{0.48\textwidth}{@{\extracolsep{\fill}}llll}
    \hline
    &\textrm{Dimer}&Infinite Chain&Experiment\\
    \hline
 $J$ (meV)& 17.7 & 19.75&18\\
 $\beta$& 0.03 & 0.05&0.09\\

    \hline
  \end{tabular*}
\end{table}

\section{Control over the singlet-triplet transition}
Our first two proposals are based on the assumption that the exchange constant $J_{1N}$ can be modified by manipulating 
the distance between the ends of the GT chains. To achieve realistic values for $J^c_{1N}$ we select horseshoe-like chains of 
lengths in the range $N\in\,[7,15]$. We have chosen to demonstrate numerically our proposals for a $N=11$ chain, because in
this case the singlet-triplet splitting is of the order of a few meV, so that it should be measurable by spectroscopic STM methods \cite{mishra21,fasel16,oteyza20}.
We have hence checked whether we can increase $J_{1N}$ by bringing the ends of the horseshoe chain 
sufficiently close. We have found that $J_{1N}\sim J_{1N}^c$ requires forces of the order of one hundred meV/\AA, which may be realized 
in STM experiments \cite{oteyza20,mishra21}.

\begin{figure}[h]
    \centering
    \includegraphics[width=\columnwidth]{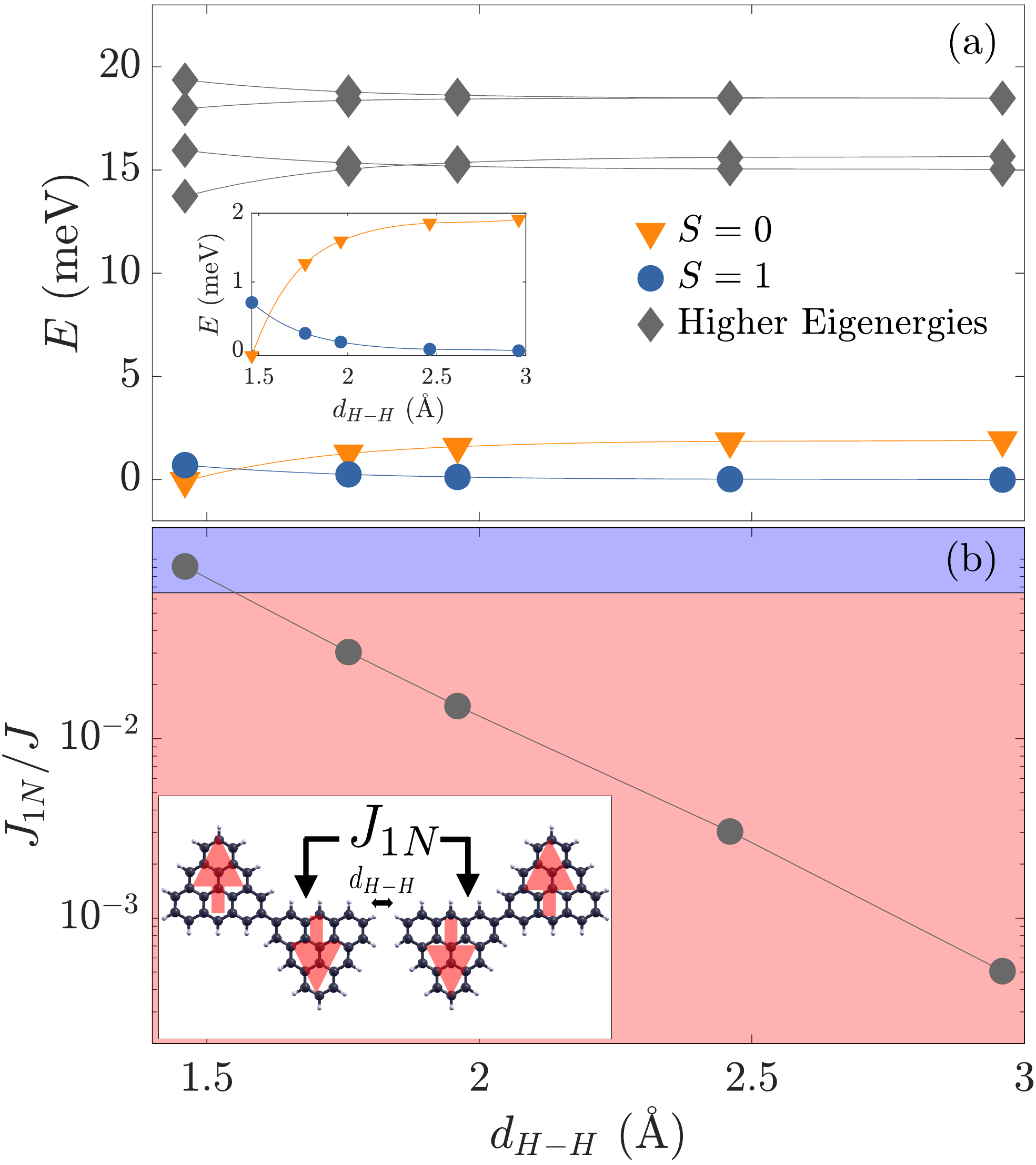}
    \caption{(a) Lowest-lying energy states as a function of $d_{H-H}$ for a $N=11$ chain. The figure shows that the singlet and triplet edge
    states lie inside the Haldane gap. Inset: Blow-up of the energy axis to highlight the singlet-triplet energy splitting.  
    (b) $J_{1N}/J$ as a function of $d_{H-H}$. The horizontal line identifies the critical $J_{1N}$ separating the regions 
    where the ground state is a triplet (pink) or a singlet (violet). Inset: sketch of the horseshoe chain end where
    the distance $d_{H-H}$ is defined. }
    \label{Figure: fig3}
\end{figure}

Within the first proposal, we assume that the distance $d_{H-H}$ between the closest hydrogen atoms at the two
chain ends can be changed in a controlled way (see the inset in Fig. \ref{Figure: fig3} for a graphical definition of $d_{H-H}$). We have 
therefore computed $J_{1N}$ as a function of $d_{H-H}$. As expected, $J_{1N}$ increases 
exponentially with decreasing $d_{H-H}$. Consequently, the energy difference between the triplet and singlet states decreases 
and, as seen in Fig. \ref{Figure: fig3}(b), the $N=11$ horseshoe GT chain experiences a singlet-triplet level
crossing  at $d_{H-H}\sim 1.6$\,\AA. The force needed to bring the two dimers in Fig. \ref{Figure: fig3} to 
$d_{H-H}\sim 1.6$\,\AA\ is of about 0.1 eV/\AA. To achieve a splitting 
$\Delta E_{ST}$ of about $1$ meV, we need to reduce the distance further to $d_{H-H}\sim 1.45$\,\AA, which require
the application of higher forces of about 0.5 eV/\AA. This is a lower bound to the full required force that does not take
into account the tensile stress caused by the deformation inside the full horseshoe chain.  However, we expect that this
contribution should not dominate for chains as long as $N=11$.
Linking atoms such as nitrogen, sulfur, phosphorous or oxygen to metalic surfaces or to graphene edges is routinely 
done in areas such as molecular electronics \cite{Cuevas2017}.\\

\begin{figure}[h]
    \centering
    \includegraphics[width=\columnwidth]{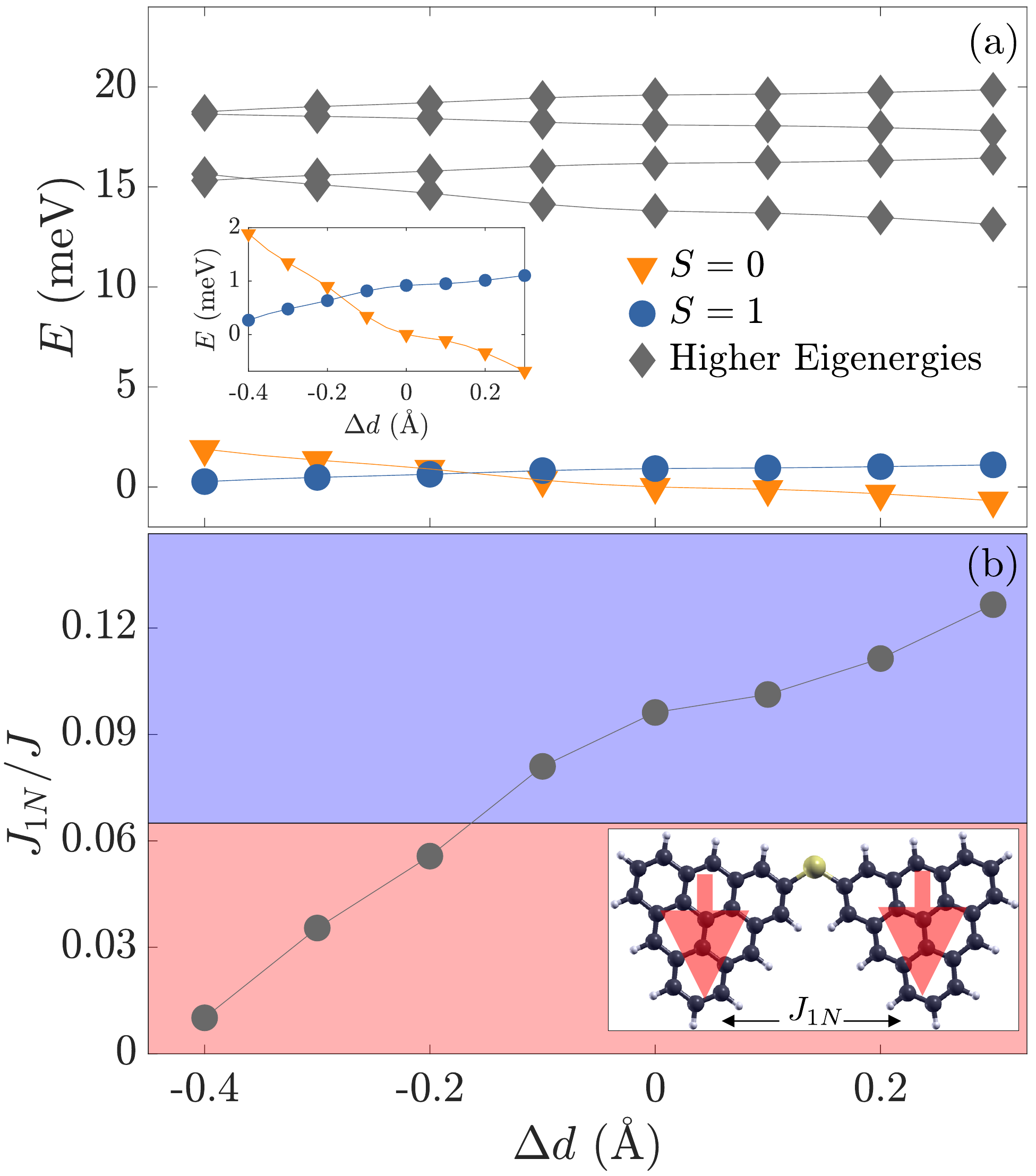}
    \caption{Same as in Figure (\ref{Figure: fig3}), where the horseshoe chain is linked by a sulfur atom.}
    \label{Figure: fig4}
\end{figure}

Our second proposal is similar to the first one but now both closest 
hydrogen atoms at the GT ends are replaced 
by a single sulfur atom that links the ends of the chain as illustrated in the inset of Fig. \ref{Figure: fig4}. 
Our DFT simulations show that the sulfur atom does not change the magnetic moment of the two adjacent GTs. We then compute 
$J_{1N}$ as a function of the distance $\Delta d$ relative to the equilibrium distance of the two edges. Our results, 
depicted in Fig.~\ref{Figure: fig4}(b), demonstrate that the singlet-triplet level crossing can be triggered by closing 
the structure by about 0.3\,\AA. We find forces now of order 0.7 eV/\AA\ for $\Delta d\sim -0.4$\,\AA, where 
$\Delta E_{ST}\,\sim\,1.5$ meV.\\

\begin{figure}[h]
    \centering
    \includegraphics[width=\columnwidth]{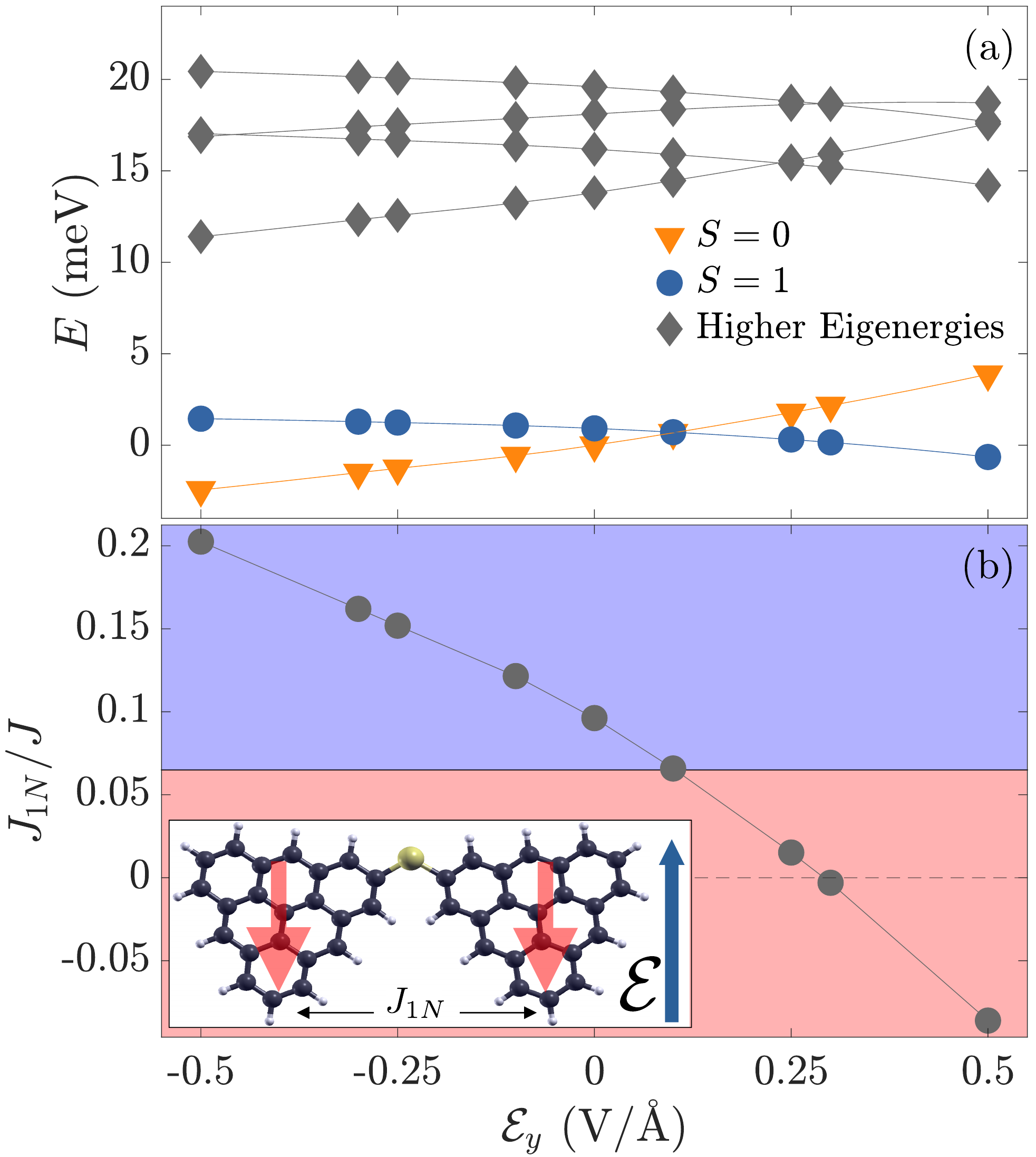}
    \caption{Same as in Figure (\ref{Figure: fig3}), where in addition to a linking sulfur atom, an electric field 
    $\bm{\mathcal E}$ is being applied along the chain main axis. }
    \label{Figure: fig5}
\end{figure}

The third proposal is based on the observation that the sulfur atom introduces an electric dipole at the chain weak link 
that renders $J_{1N}$ susceptible to an external electric field $\bm{\mathcal E}$.
Our first-principles simulations confirm that this is the case, $\bm{\mathcal E}$ being most effective when pointing
along the symmetry axis of the horseshoe-shaped GT chain. We plot in Fig.~\ref{Figure: fig5} 
the energies of the singlet 
and triplet states, as well as the exchange constant $J_{1N}$ as a function of ${\mathcal  E}_y$. Indeed, we find that there 
is a level crossing from singlet to triplet at about ${\mathcal E}_y\sim0.1$ V/\AA. We find a threshold coupling 
$J^c_{1N} \sim 0.06\, J$ that agrees well with the estimates from our exact diagonalization studies for $N=11$ and $\beta=0.05$, 
shown in Fig. \ref{Jcritic}(a). Our numerical study provides strong evidence that the single-triplet transition can be 
sensitively controlled by an electric field.
To clarify the physical origin of the electric-field based mechanism, we have first checked that the spin of each GT 
remains equal to 1 and that the exchange constant $J$ between the GTs in the chain changes only by about 1-2 meV 
even for the largest simulated fields. We have also checked that the electric field does not affect the chain geometry 
even around the sulfur atom. We have then investigated the redistribution of atomic Mulliken charges at 
the terminating GTs when the sulfur atom is present at the junction. We have then found that the combined influence of the sulfur atom and the electric field induces 
an internal dipole in those GTs. Further calculations in \cite{supmat} show that $J_{1N}$ is linearly proportional not only to the 
external electric field but also to the electric dipole moment at the junction.

\section{Conclusions}
 We have demonstrated that GT chains are faithful realizations of the nearest neighbor spin-1 AFM
chain by combining spin-model simulations with first-principles calculations. We have advanced three proposals 
for experiments that may trigger and control the singlet triplet transition of odd-numbered chains, thus opening
the door for their future use as quantum devices. Our calculations indicate that the application of an external 
electric field is particularly feasible.

\section*{Acknowledgements}
G. M.-C., A. G.-F. and J. F. have been funded by Ministerio de Ciencia, Innovación y Universidades, Agencia Estatal de 
Investigación, Fondo Europeo de Desarrollo Regional via the Grant PGC2018-094783. G. M.-C. has also been supported by Programa 
``Severo Ochoa'' de Ayudas para la investigación y docencia del Principado de Asturias. G.M.-C. acknowledges the financial 
support and hospitality of the Wigner Research Centre for Physics. This project has been supported by Asturias FICYT under grant AYUD/2021/51185 with the support of FEDER funds. L. O. also acknowledges support of the
National Research, Development and Innovation (NRDI) Office of Hungary and the Hungarian Academy of Sciences through the Bolyai 
and Bolyai+ scholarships. This research is supported by the NRDI Office within
the Quantum Information National Laboratory of Hungary and  through the research grants K131938 and FK124723.

\bibliography{biblio}

\end{document}


\lstset{frame=single,language=Matlab,aboveskip=3mm,belowskip=3mm,showstringspaces=false,columns=flexible,basicstyle={\small\ttfamily},numbers=left,stepnumber=1,numberstyle={\tiny\color{gris}},keywordstyle=\color{blue},commentstyle=\color{black},stringstyle=\color{black},breaklines=true,breakatwhitespace=true,tabsize=3}

\maketitle
\appendix
\section{Derivatives BLBQ Hamiltonian}

We start from the classical BLBQ Hamiltonian:
\begin{equation}
H=\dfrac{1}{2}\sum\limits_{i\neq j}^NJ_{ij}\left[\mathbf{S}_i\cdot \mathbf{S}_j+\beta_{ij}\left(\mathbf{S}_i\cdot \mathbf{S}_j\right)^2\right].
    \label{BLBQ}
\end{equation}
The first derivate of Eq. \ref{BLBQ} respect to $\mathbf{S}_m$ is given by
\begin{equation}
\begin{split}
    &\dfrac{\partial H}{\partial \mathbf{S}_m}=\dfrac{1}{2}\sum\limits_{i\neq j}^NJ_{ij}\left(\delta_{im}\mathbf{S}_j+\delta_{jm}\mathbf{S}_i+2\beta_{ij}\left(\mathbf{S}_i\cdot \mathbf{S}_j\right)\left[\delta_{im}\mathbf{S}_j+\delta_{jm}\mathbf{S}_i\right]\right)=\\&=\sum\limits_{j(j\neq m)}J_{jm}\left[\mathbf{S}_j+2\beta_{jm}\left(\mathbf{S}_m\cdot\mathbf{S}_j\right)\mathbf{S}_j\right],
    \end{split}
\end{equation}
and the second derivative is given by
\begin{equation}
\begin{split}
    &\dfrac{\partial^2 H}{\partial \mathbf{S}_m\partial \mathbf{S}_n}=J_{mn}\left(1+2\beta_{mn}\left[\left(\mathbf{S}_m\cdot\mathbf{S}_n\right)+\mathbf{S}_n\circ\mathbf{S}_m\right]\right),
    \end{split}
\end{equation}

where $\circ$ is the dyadic product. Notice that the interaction energy between $\mathbf{S}_n$ and $\mathbf{S}_m$ is given by
\begin{equation}
\begin{split}
    &\delta E^{(2)}_{nm}=\delta\mathbf{S}_m\dfrac{\partial^2 H}{\partial \mathbf{S}_m\partial \mathbf{S}_n}\delta\mathbf{S}_n=\\&=J_{nm}\left(\left(\delta\mathbf{S}_m\cdot\delta\mathbf{S}_n\right)+2\beta_{nm}\left[\left(\delta\mathbf{S}_m\cdot\delta\mathbf{S}_n\right)\left(\mathbf{S}_m\cdot\mathbf{S}_n\right)+\left(\delta\mathbf{S}_m\cdot\mathbf{S}_n\right)\left(\mathbf{S}_m\cdot\delta\mathbf{S}_n\right)\right]\right),
    \end{split}
\end{equation}
where $\delta\mathbf{S}_m$ is the infinitesimal variation of $\mathbf{S}_m$. In a collinear model ($\mathbf{S}_n\parallel\mathbf{S}_m$) we get that for transversal spin fluctuations $\delta\mathbf{S}_n\cdot\mathbf{S}_m=\mathbf{S}_n\cdot\delta\mathbf{S}_m=0$ it is obtained  
\begin{equation}
    \delta E_{nm}^{(2)}=J_{nm}\left[1+2\beta_{nm}\left(\mathbf{S}_n\cdot\mathbf{S}_m\right)\right]\delta\mathbf{S}_n\cdot\delta\mathbf{S}_m.
\end{equation}
\newpage
\section{Singlet-Triplet Splitting}
\begin{figure}[h]
    \centering
    \includegraphics[width=\columnwidth]{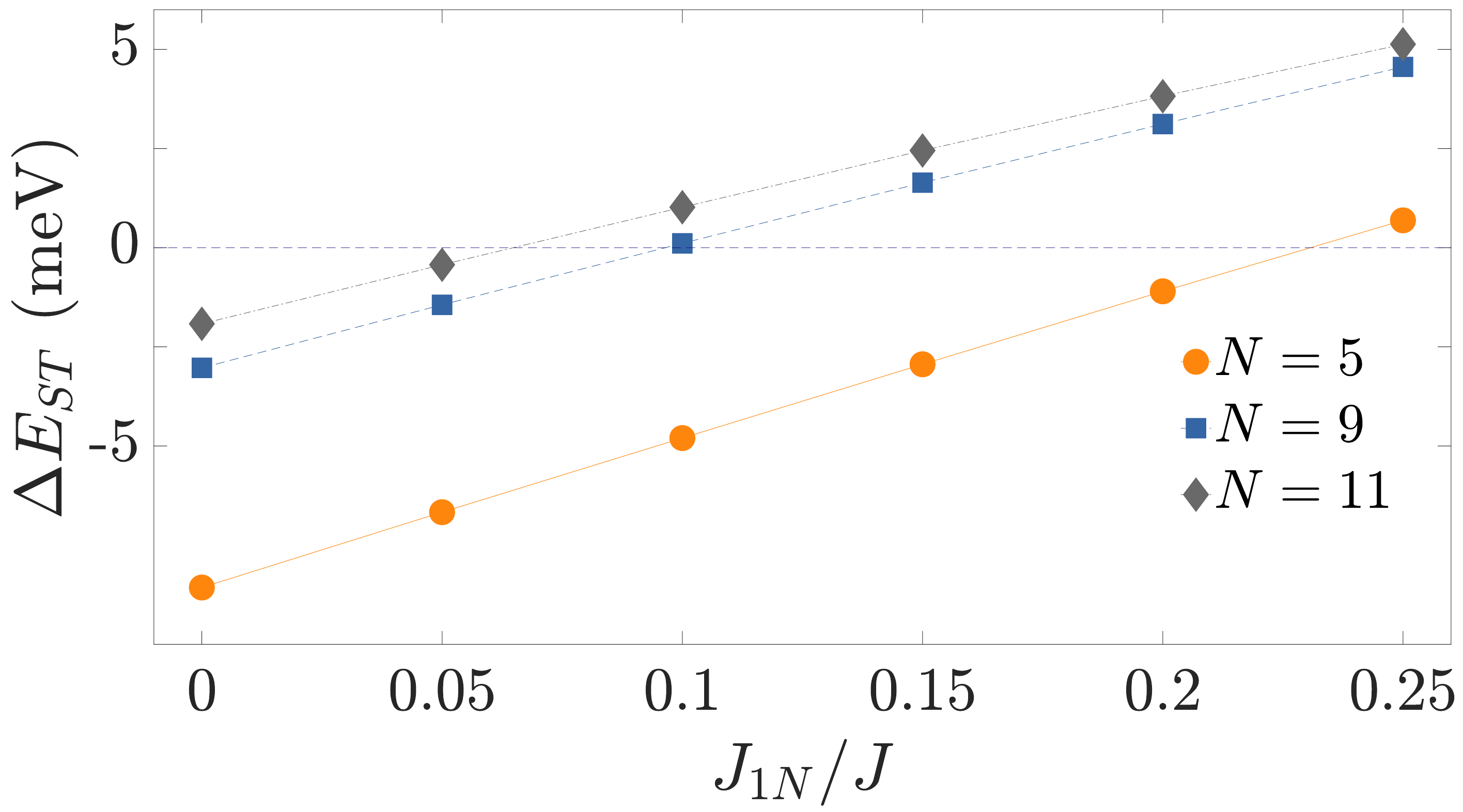}
    \caption{Singlet triplet energy-splitting as a function $J_{1N} /J$
for $N=$ 5, 9 and 11 chain lengths, all having $J = 19.75$ meV
and $\beta = 0.05$.}
    \label{supplemental}
\end{figure}

The energy splitting $\Delta E_{ST}=E(S=1)-E(S=0)$ of the singlet and triplet states lying deep in the Haldane gap can be obtained by quantizing the spin Hamiltonian \eqref{BLBQ}. As we argue in the main text introducing a coupling $J_{1N}$ between the spins on the edge of the chain a transition between the singlet and triplet state can be evoked. In Fig. \ref{supplemental} we show the energy splitting $\Delta E_{ST}$ in terms of $J_{1N}$ for various chain lengths. It can be observed that the splitting is linear in the coupling. Also it is evident that longer chains require weaker coupling in order to drive them through the singlet-triplet transition.
The energy scale of the singlet/triplet splitting can be tuned in the order of a couple of meVs, i. e., the splitting is easily resolved in experiments.
\vspace{1cm}
\newpage
\section{Electric Dipole on the Juntion}
\begin{figure}[h]
    \centering
    \includegraphics[width=\columnwidth]{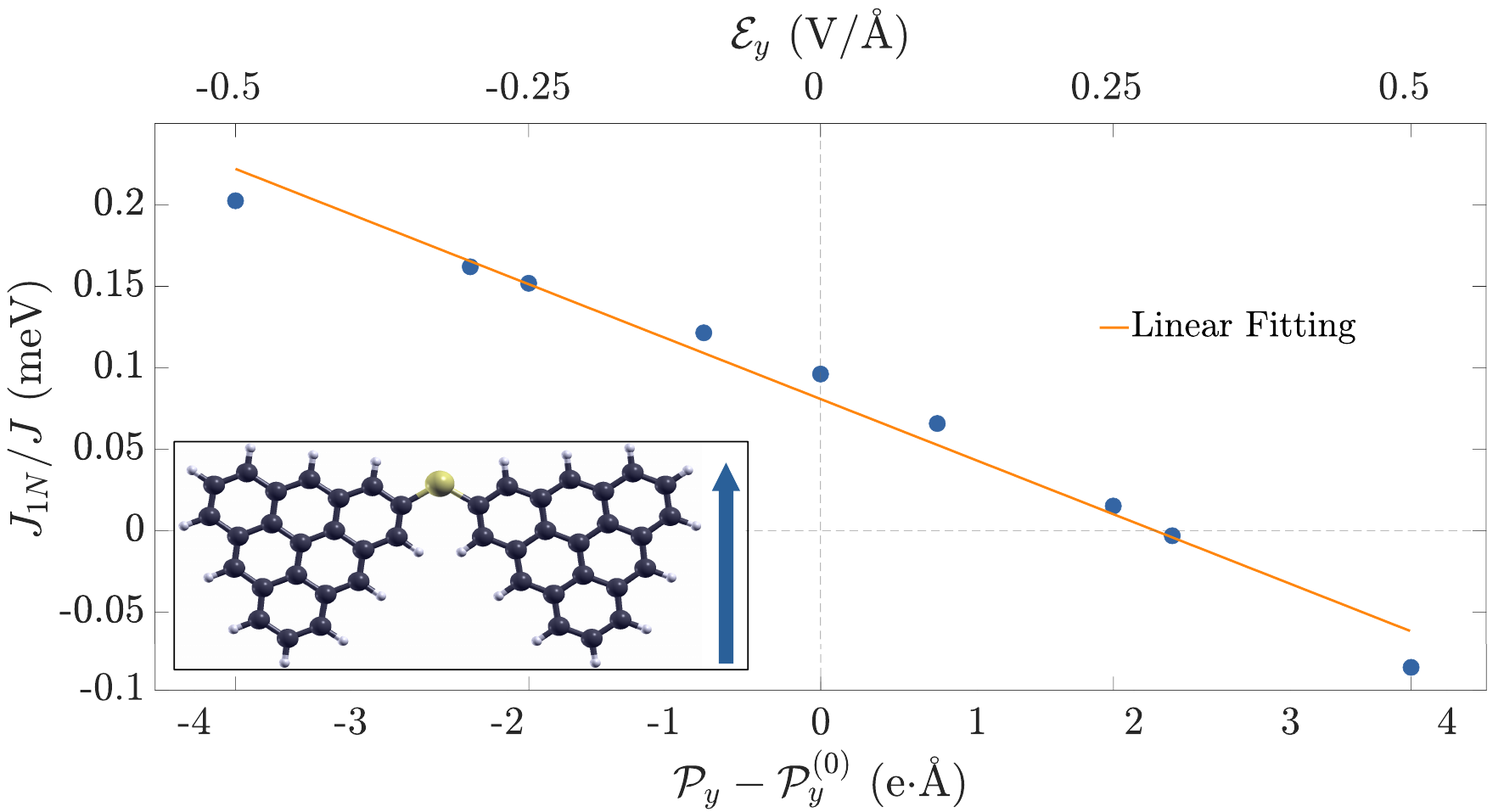}
    \caption{$J_{1N}$ parameter as a function of the applied electric field and the total electric dipole on the junction. }
    \label{supplemental2}
\end{figure}

In the main text we state that that the electric tunability of the coupling constant $J_{1N}$ is proportional to the dipole moment assotiated with the junction. In Fig. \ref{supplemental2} we corroborate this statement by calculating the coupling constant in terms of the applied electric field and the calculated electric dipole.
As it can be observed there is, to a good approximation, a linear relation between $J_{1N}$ and the induced dipole moment $\mathcal{P}_y$.
Performing a linear fit to the data points as function of $\mathcal{P}_y-\mathcal{P}_y^{(0)}$ where $\mathcal{P}_y^{(0)}$ is the electric dipole at zero external field, we obtain $J_{1N}/J=-0.038$ ($e\cdot$\AA)$^{-1}\left( \mathcal{P}_y-\mathcal{P}_y^{(0)} \right)+0.081$.


